\documentstyle[prb,eqsecnum,aps]{revtex}

\begin{document}
\draft
%\preprint{HEP/123-qed}
\title
{
Ground state of the spin-$\frac{1}{2}$ Heisenberg
antiferromagnet on a two-dimensional square-hexagonal-dodecagonal lattice
}
\author
{
Piotr Tomczak
}
\address
{
Magnetism Theory Division, Physics Department,
Adam Mickiewicz University,\protect\\
Umultowska 85, 61-614 Pozna\'n, Poland
}
\author
{
Johannes Richter
}
\address
{
Institut f\"ur Theoretische Physik, Otto-von-Guericke Universit\"at Magdeburg,\protect\\
P.O.B. 4120, 39016 Magdeburg, Germany
}
%\date{\today}
\maketitle
\begin{abstract}
Up to now, the existence of the the magnetic
N\'eel Long Range Order (NLRO) in nearest neighbor, spin-$\frac{1}{2}$
antiferromagnetic (AF) lattice systems has been examined
for seven, from the eleven existing, two-dimensional, uniform lattices.
Plaquettes forming these
uniform (Archimedean) lattices (e.g. square, triangular, kagom\'e)
are different regular polygons.
An investigation of the NLRO in the ground
state of AF spin systems on the seventh uniform (bipartite)
lattice consisting
of squares, hexagons and dodecagons is presented.
The NLRO is shown to occur in this system.
A simple conjecture concerning the existence of the NLRO
in the ground state of antiferromagnetic, spin-$\frac{1}{2}$ systems
on two dimensional, Archimedean lattices, is formulated.
\end{abstract}


%\narrowtext

\section{INTRODUCTION}
Despite of a great progress, stimulated originally by the
discovery of high-$T_c$ superconductivity, made
in last years in the understanding of the nature of the ground
state of quantum Heisenberg
antiferromagnets for low values of spin variables on two-dimensional
lattices, the problem is far from being solved. The basic question,
if there exists the N\'eel Long Range Order (NLRO) in the ground state
of an antiferromagnetic spin-$\frac{1}{2}$ system on a given
lattice is still not completely answered.
In fact, a subtle interplay between quantum effects and lattice
effects, rather difficult to deal with, may lead
to nontrivial properties of the ground state.
Restrict now the attention to the most simple case of spin
system with equal, antiferromagnetic, nearest neighbor interactions:
\begin{equation}
\label{eq:1}
      H =  \sum _{<i,j>}\roarrow{S}_i \cdot \roarrow{S}_j.
\end{equation}
The natural way to represent this kind of spin system is to
put the spins onto vortices of a so-called uniform (Archimedean) lattice
and to assign the antiferromagnetic spin-spin interactions along
the lines connecting the nearest spins.
{\it Uniform} means that all polygons of the smallest area
 built from nearest neighbor interactions (plaquettes) are
regular, and consequently, each lattice site has the same
local environment. For example, in the case of
a kagom\'e lattice
there exist two kinds of such polygons: triangles and hexagons.
Up to now, the problem of the eventual existence of the NLRO in
a spin-$\frac{1}{2}$ system residing on a uniform lattice has been
investigated for seven uniform lattices only. Let us summarize
the results of those investigations. First, take into account {\it
unfrustrated}
lattices. In the case of square\cite{Man}, honeycomb\cite{MaFr} and
CaVO\cite{Wei} lattices there exist a well grounded opinion
that the ground
state is N\'eel-like ordered. For {\it frustrated} lattices, however,
the situation is more complicated. There exist three sublattice
magnetic order in spin-$\frac{1}{2}$ system on the triangular
lattice\cite{Deu,Els,LeBe,Bo}
and probably does not exist in the case of kagom\'e lattice\cite{Sin}.
For zig-zag-ladder lattice\cite{Nor} it seems to exist an
incommensurate spiral order. The most complicated situation
is for the model examined by Shastry and Sutherland\cite{SaSu}
and more recently by Albrecht and Mila\cite{AlMi} which
corresponds to a Heisenberg antiferromagnet on an uniform lattice
for $J_1=J_2$ (notation taken from Ref. [11]). In that case
spin wave theory gives a disordered ground state, whereas
Schwinger-boson
mean field theory predicts NLRO. However, for $J_2/J_1\approx1.1$
NLRO becomes unstable and gives way to spiral order\cite{AlMi}.

We formulate now the simple conjecture which seems to apply to
the problem of the NLRO in the ground state of spin-$\frac{1}{2}$
systems with nearest neighbor interactions on uniform lattices.
At first, however, let us classify all investigated so far
uniform lattices according to
the number of edges of polygons of the smallest area 
bound by the nearest neighbor interactions.
There exist three possible cases: {\it i}) only even polygons
(i.e., with even number of edges) are present, e.g.,
square lattice - squares, honeycomb lattice - hexagons,
CaVO lattice\cite{Wei} - squares and octagons.
Call those lattices the even ones.
{\it ii}) only odd polygons are present, triangular lattice -
triangles. This is the only one odd lattice and
{\it iii}) even and odd polygons are present, e.g.,
kagom\'e lattice - hexagons and triangles, zig-zag-ladder
lattice\cite{Nor} and Shastry - Sutherland lattice\cite{SaSu}
- squares and triangles. Those are even-odd lattices.
The above classification allows one to formulate the following
conjecture: the antiferromagnetic spin-$\frac{1}{2}$ system on
an even-odd lattice has no NLRO in its ground state. If the lattice
is even {\it or} odd - the ground state is N\'eel-like ordered.

The aim of this paper is to report an additional example which
 does support
this conjecture. This example concerns the spin-$\frac{1}{2}$
system located on the last possible, yet not investigated, even
Square - Hexagonal - Dodecagonal (SHD) lattice, see Fig. 1.
Note that there exist only four
even, one odd, and six even-odd uniform (Archimedean) lattices.

\section{METHOD}
To answer the question about the NLRO in the ground state of the
spin system on the {\it bipartite} lattice from Fig. 1 a variational
approach
developed by Huse and Elser\cite{HuEl} and improved later to restore
the $S^x, S^y$ symmetry by Carneiro, Kong and Swendsen\cite{Car}
was applied.
It was also shown that within this method one can obtain reliable
results for both magnetically N\'eel ordered and disordered
systems\cite{HuEl,Tom}.
Let us remind the basic points of this method.
At the beginning, the variational wave function
is expanded into the complete set of Ising states
$\mid\!\alpha\rangle$ in the subspace of total $S^z=0 $:
\begin{equation}
\label{eq:2}
\mid\!\!\Psi\rangle =\sum_{\alpha}\,\exp(\frac{1}{2}\tilde{H})\mid\!\alpha\rangle ,
\end{equation}
and the operator $\tilde{H}$, diagonal in that base is defined as
\begin{equation}
\label{eq:3}
\tilde{H} = 2 \pi i (\sum _{i\in B}\frac{1}{2} - S^{z}_{i}) +
\sum _{i,j}K(r_{ij}) S^{z}_iS^{z}_j.
\end{equation}
The first term  produces a proper sign (phase) for a given
state $\mid\!\alpha\rangle$ (the first sum runs over
spins belonging to the $B$ sublattice) whereas the second one, taking into
account in a variational way the long-range correlations in the spin system,
gives the value of the amplitude for that state.
The second sum runs over all possible pairs of spins. The next step of this
method is to find a minimum of the ground state energy
$\langle \Psi\! \mid\!H\! \mid\! \Psi \rangle / \langle\Psi\! \mid\! \Psi \rangle $
with respect to parameters $K(r_{ij})$ and, subsequently, to calculate
for those $K(r_{ij})$
the expectation values of the operators which  characterize the LRO
in the ground state of a given spin system. In the case of small clusters
this can be accomplished by taking into account the whole subspace of Ising
states, in the case of larger ones - within a Monte-Carlo
approach\cite{HuEl,Car}.
Finally, those values are extrapolated to the thermodynamic limit.

Actually, two expectation values of such operators were calculated:
the squared sublattice magnetization and, which is new within this
approach, the spin gap, i.e., $E_1-E_0$, where $E_0$ and $E_1$ stand
for the ground state energy (sector $S^z=0$) and the energy
with one flipped spin (sector $S^z=1$), respectively.

How to choose the variational parameters $K(r_{ij})$ in Eq. (3) for
the bipartite SHD lattice? Several choices of the
parameter space were tested in variational calculations
done for spin systems on finite clusters of other bipartite lattices:
for 18 spins on the linear chain and on the honeycomb lattice,
for 16 spins on the CaVO lattice and for 12 spins on the SHD lattice.
The whole $S^z=0$ basis in expansion (2) was taken into account.
The minimal value of the variance of the ground state energy has
been obtained in all cases in three parameter space:
$(K_{AA}, K_{AB}, \sigma)$. It means that $K(r_{ij})=K_{AA}/r^{\sigma}$
when spins at the distance $r_{ij}$ belong to the same sublattice,
$K(r_{ij})=K_{AB}/r^{\sigma}$ otherwise,
and $r_{ij}$ is the Manhattan metric (the shortest path over bonds),
instead of the commonly used Euclidean one.
Henceforth all expectation values of operators are calculated
for the latter choice of the variational parameters.

\section{RESULTS}
To estimate the quality of this approach to the problem, the
ground-state energy, the squared magnetization, the correlation
function for the highest distance in the cluster and the spin
gap were calculated by the direct diagonalization and, in addition, the variational
approach described above. The results for some clusters
on different bipartite lattices are collected and compared in
Table I. Notice that for all clusters considered in this paper
periodic boundary conditions were used.
In the variational approach all the base functions of the $S^z=0$
subspace were taken into account. From Table I one can see that this
variational method overestimates a little bit the tendency
towards LRO -
the variational values $m^2$ and the correlations are bigger
than the exact ones and the variational values of the spin gap
are smaller than the exact ones.

Now, let us take into account larger clusters. The variational
results for 48-, 108-, and 192-spin clusters are collected in the
Table II and the finite size analysis is presented in Figs. 2 - 5.
It has also been checked that the energy minimum for the lattice of $108$
spins is attained for values of the variational parameters $K_{AA}=-2.77$,
$K_{AB}=-2.85$,
and  $\sigma=1.17$, different but very slightly from those for $48$
spins, so it was decided to keep them unchanged in the calculations 
for 192-spin cluster.

One can assume that the leading term of the finite-size correction
of the ground state
energy per bond $E$ resulting from the cutoff of the long-wavelength
magnons which are linear in $k$, is, like in the case of other
translationally invariant systems, $N^{-3/2}$. The data from
Table II can be fitted to this dependence and hence the energy
per bond, $E_\infty$ in the thermodynamic limit is obtained:
$E(N)=E_\infty+aN^{-3/2}$ with $E_\infty=-0.3605$ and $a=-0.8200$.
This dependence is shown in Fig. 2.
Using in the fitting only the data for 48, 108 and 192 spins
one obtains $E_{\infty}=-0.3607$, and
practically the same value of $E_{\infty}$ gives fitting
to the formula $E(N)=E_{\infty}+a'N^{-3/2}+bN^{-2}$.

In Fig. 3.  the order parameter squared
as a function of the system size $N^{-1/2}$  is depicted.
The square of order parameter should scale as $N^{-1/2}$, therefore,
for small $N$, corrections of higher orders may be important.
Thus we decided to take into account only the data for $N=$48, 108 and
192 spins in the extrapolation. This leads to the following
form for the square of sublattice magnetization as a function
of N: $m^2(N)=m^2_\infty+cN^{-1/2}$ with
$m^2_\infty=0.1007$ and $c=0.5058$. One can conclude
that the long-range magnetic order persists
in the ground state of this spin system.

Finally, an additional argument, supporting the existence
of the NLRO order, is presented in Fig. 4 which shows
the extrapolation of the spin gap to the thermodynamic
limit according the relation $\Delta(N)=\Delta_\infty+dN^{-1}$
with $\Delta_\infty=-0.0260$ and $d=6.514$. Note that the
value of $d$ is consistent with the leading-order spin
wave result\cite{Ham} for the finite-lattice energy gap:
$\Delta(N)=2zN^{-1}+\ldots$, where $z$ is the number of nearest
neighbors. Keeping in mind
the tendency of the variational method to the underestimation
of the spin gap one can regard the ground state as being
gapless.

\section{SUMMARY}
In this paper, the first investigation of the
ground state of the Heisenberg spin-$\frac{1}{2}$ system
on square-hexagonal-dodecagonal lattice is presented.
The calculated valus of $m^2_\infty$ and $\Delta_{\infty}$ seem to
be an evidence for the existence of two-sublattice N\'eel long-range
magnetic order in this system. This result can also be
regarded as an argument supporting
a very simple conjecture concerning the existence
of the magnetic NLRO in the ground state of the Heisenberg
antiferromagnets on uniform lattices, however further studies,
especially for even-odd lattices,
supporting (or not) this conjecture are required.\\

{\bf Acknowledgement}

We thank Professor Ryszard Ferchmin for helpful discussions
and reading the manuscript.
One of us (P.T.) thanks the
Otto-von-Guericke University  for support. We acknowlege support
from the Polish Committee for Scientific Reaserch (Project No. 2 PO3B 046 14)
and from the Deutsche Forschungsgemeinschaft (Project No. Ri 615/1-2).
Some of the calculations were performed at the Pozna\'n Supercomputer and
Networking Centre under grant No. 8T11F 010 08p04.

\begin{figure}
\label{pierw}
\caption
{
SHD (Square-Hexagonal-Dodecagonal) lattice. The fourth uniform,
even, bipartite lattice. Sublattices are marked by circles
and squares. Spins residing on A-sublattice interact only with
its nearest neighbors from B-sublattice. The 48-spin cluster
used in calculations is also marked. Other clusters have the same shape.
}
\end{figure}

\begin{figure}
\label{drug}
\caption
{
Energy per bond for the spin system on the SHD lattice
 as a function of $N^{-3/2}$.
}
\end{figure}

\begin{figure}
\label{czwa}
\caption
{
Plot of the $m^2$ as a function
 of $N^{-1/2}$.
Squares - values obtained by applying the variational
 method. Straight line -  fit to the squares.
Sizes of squares are comparable to the error bars.}
\end{figure}

\begin{figure}
\label{piat}
\caption
{
Plot of the spin-gap $\Delta=E_1-E_0$ vs $1/N$.
Errors result from adding the errors for $E_0$ and $E_1$.
}
\end{figure}

\begin{table}
\caption{Comparison between exact and variational results for finite
clusters for different bipartite lattices. All the variational results
were obtained in three parameter space $K_{AA}, K_{AB}, \sigma$.}
\begin{tabular}{cccccc}
% &\multicolumn{2}{c}{$D_{4h}^1$}&\multicolumn{2}{c}{$D_{4h}^5$}\\
     &    &  linear chain  &  honeycomb lattice  & CaVO lattice & SHD lattice\\
     &    &18-spin cluster&18-spin cluster&16-spin cluster&12-spin cluster\\
\tableline
Ground state&exact &-0.4457&-0.3740&-0.3800&-0.3850\\
energy, per bond&variational&-0.4450&-0.3698&-0.3733&-0.3803\\ \tableline
$m^2$&exact&0.1851&0.2481&0.2504&0.2913\\
     &variational&0.1859&0.2509&0.2616&0.2979\\ \tableline
correlation for largest&exact&0.0859&0.1655&0.1468&0.1924\\
distance in cluster&variational&0.0752&0.1714&0.1724&0.2032\\ \tableline
   gap&exact& 0.241 & 0.394 & 0.446 & 0.568 \\
 &variational& 0.241 & 0.323 & 0.378 & 0.519\\
 \end{tabular}
 \label{table1}
 \end{table}

\begin{table}
\caption{The ground state energy, the squared sublattice magnetization,
the energy of the first excitation $E_1$
and the spin gap for some clusters on SHD lattice. In a case of 12 spin
cluster the values were obtained in the whole $S^z=0$ sector, for bigger
clusters the Monte-Carlo method was applied. Statistical errors,
in parantheses, are the last digit.}
\begin{tabular}{ccccc}
 N  &  $E_0/bond$  &  $ m^2$  & $E_1$ & $E_1-E_0$ \\
\tableline
12  &-0.3803   &0.2979 & -6.3285   &0.519\\ \tableline
48  &-0.3628(4)&0.1743(8)& -26.014$\pm0.019$ &0.107$\pm$0.038\\ \tableline
108 &-0.3614(2)&0.1476(6)& -58.506$\pm0.034$ &0.042$\pm$0.060\\ \tableline
192 &-0.3609(2)&0.1384(5)&-103.948$\pm0.019$ &0.003$\pm$0.080\\
 \end{tabular}
 \label{table2}
 \end{table}

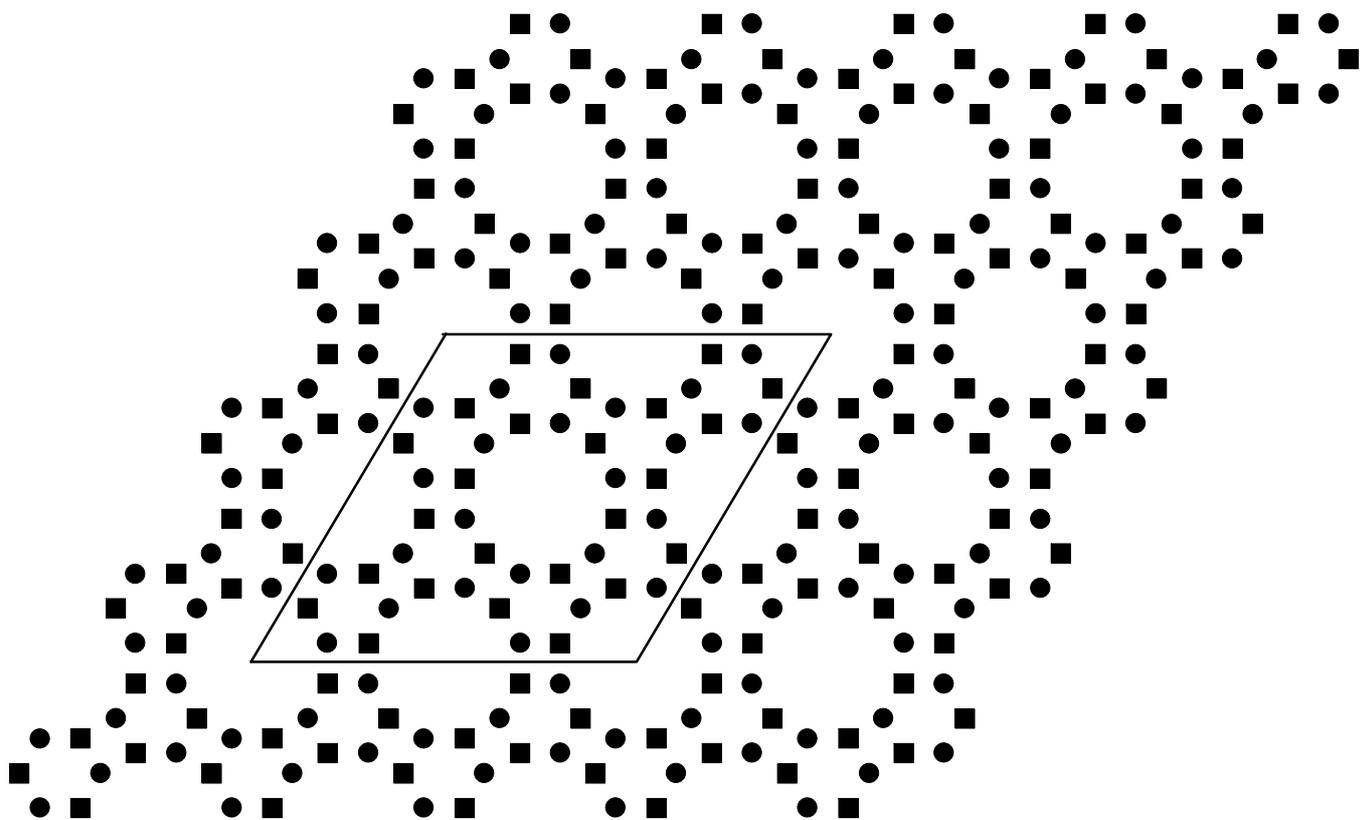
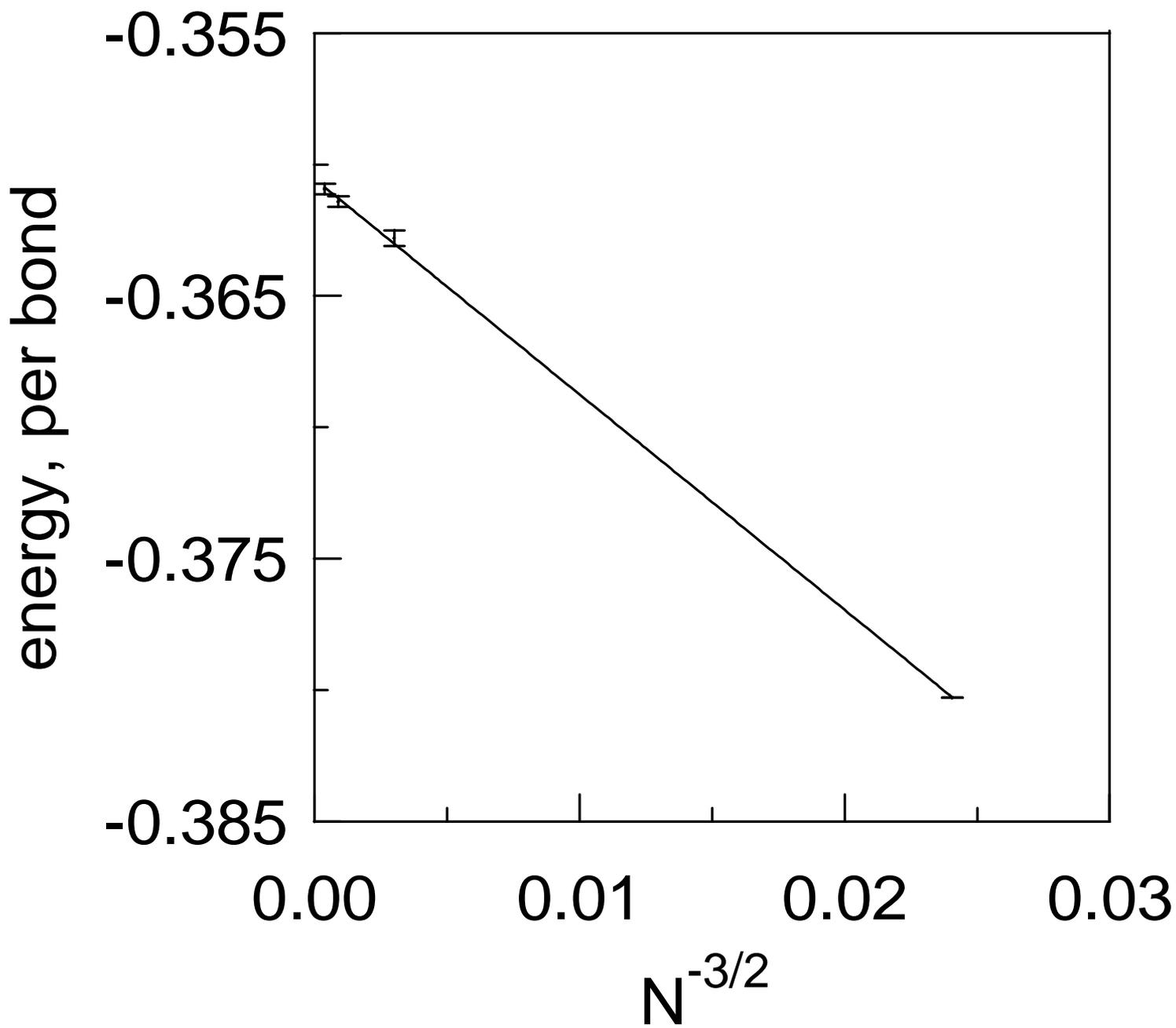
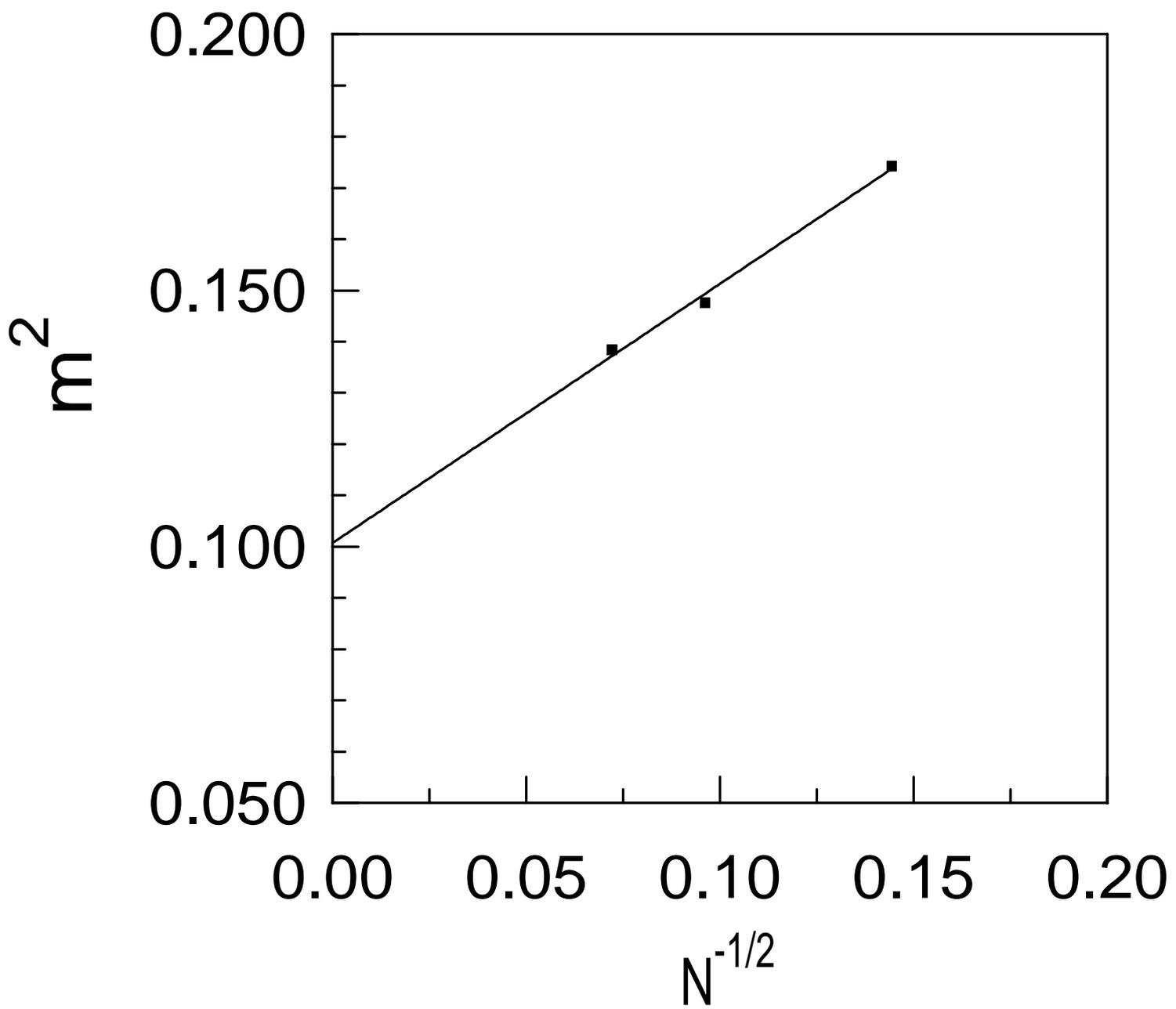
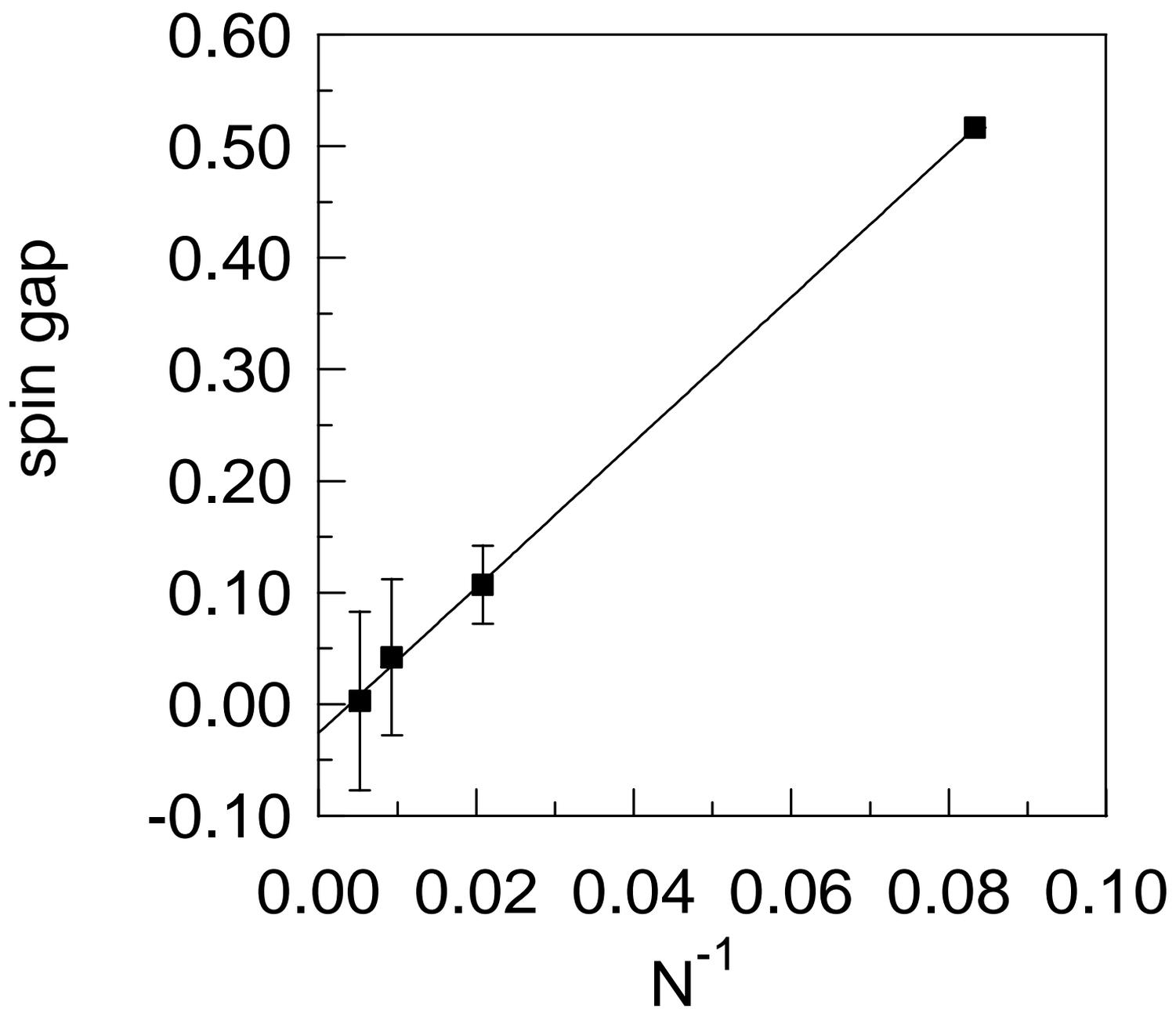
\end{document}